\documentclass[11pt]{article}

\usepackage{amssymb,amsmath}
\usepackage[noend]{algpseudocode}
\usepackage{graphicx}
\usepackage{verbatim}
\usepackage{url,xspace,hyperref}
\usepackage{cite}
\usepackage{caption}
\usepackage{subcaption}
\usepackage{multirow}
\usepackage{multicol}
\usepackage{theorem,latexsym}
\usepackage{amsmath,amssymb,enumerate}
\usepackage{xspace}
\usepackage{algorithm,algorithmicx}
\usepackage{float}

\usepackage{fullpage}

\usepackage{geometry}
\newtheorem{theorem}{Theorem}[section]
\newtheorem{proposition}[theorem]{Proposition}
\newtheorem{lemma}[theorem]{Lemma}

\newtheorem{remark}[theorem]{Remark}
\newtheorem{fact}[theorem]{Fact}

\def\sse{\subseteq}

\newcommand{\R}{\mathbb{R}}

\newcommand{\dto}{\ensuremath{\mathsf{DTO}}\xspace}
\newcommand{\sto}{\ensuremath{\mathsf{STO}}\xspace}

\newcommand{\ignore}[1]{}

\newcommand{\qedsymbol}{\rule{0.6em}{0.6em}}

\title{Quasi-Polynomial  Algorithms for Submodular Tree Orienteering and Other Directed Network Design Problems}
\author{Rohan Ghuge\thanks{Industrial and Operations Engineering Department, University of Michigan. Supported in part by NSF CAREER grant CCF-1750127.} \and Viswanath Nagarajan$^\star$}

\begin{document}
\maketitle

\begin{abstract}
    We consider the following general network design problem on directed graphs. The input is  an asymmetric metric $(V,c)$, root $r^{*}\in V$, monotone submodular function $f:2^V\rightarrow \R_+$ and budget $B$. The goal is to find an $r^{*}$-rooted arborescence $T$ of cost at most $B$ that maximizes $f(T)$. Our main result is a simple quasi-polynomial time $O(\frac{\log k}{\log\log k})$-approximation algorithm for this problem, where $k\le |V|$ is the number of vertices in an optimal solution. To the best of our knowledge, this is the first non-trivial approximation ratio for this problem. As a consequence we obtain an $O(\frac{\log^2 k}{\log\log k})$-approximation algorithm for  directed (polymatroid) Steiner tree in quasi-polynomial time. We also extend our main result to a setting with additional length bounds at vertices, which leads to improved $O(\frac{\log^2 k}{\log\log k})$-approximation algorithms for the single-source buy-at-bulk and priority Steiner tree problems. For the usual directed Steiner tree problem, our result  matches the best previous approximation ratio~\cite{GLL18}. Our algorithm has the advantage of being deterministic and faster:  the runtime is $\exp(O(\log n\, \log^{1+\epsilon} k))$. For polymatroid Steiner tree and single-source buy-at-bulk, our result improves prior approximation ratios  by a logarithmic factor. For directed priority Steiner tree, our result seems to be the first non-trivial approximation ratio.     All our approximation ratios are  tight (up to constant factors) for quasi-polynomial algorithms.
\end{abstract}

\thispagestyle{empty}

\newpage

\setcounter{page}{1}

\section{Introduction}
Network design problems, involving variants of the {\em minimum spanning tree} (MST) and {\em traveling salesman problem} (TSP), are extensively studied in approximation algorithms. These problems are also practically important as they appear in many applications, e.g. networking and vehicle routing. 
Designing algorithms for problems on directed networks is usually much harder than their undirected counterparts. This difference is already evident in the most basic MST problem: the undirected case admits a simple greedy algorithm whereas the directed case requires a much more complex algorithm~\cite{Ed67}. Indeed, one of the major open questions in network design  concerns the directed Steiner tree problem. Given a directed graph with edge costs and a set of terminal vertices, the goal here is a minimum cost arborescence that contains all terminals. No polynomial-time poly-logarithmic approximation is known for directed Steiner tree. This is in sharp contrast with  undirected Steiner tree, for which a $2$-approximation is folklore and there are even better constant approximation ratios~\cite{ByrkaGRS13,RobinsZ05}. 

In this paper, we consider a variant of directed Steiner tree, where the goal is to find an arborescence maximizing the number (or profit) of vertices subject to a hard constraint on its  cost. We call this  problem {\em directed tree orienteering} (\dto). To the best of our knowledge, this problem has not been studied explicitly  before. Any $\alpha$-aproximation algorithm for \dto implies an $(\alpha\cdot \ln k)$-approximation algorithm for directed Steiner tree, using a set-covering approach. No approximation preserving reduction  is known in the reverse direction: so approximation algorithms for directed Steiner tree do not  imply anything for \dto. In this paper, we obtain a {\em quasi-polynomial} time $O(\frac{\log k}{\log\log k})$-approximation algorithm for \dto, where $k$ is the number of vertices in an optimal solution. This  also implies an $O(\frac{\log^2 k}{\log\log k})$-approximation algorithm for directed Steiner tree (in quasi-polynomial time) where $k$ denotes the number of terminals. 

In contrast to \dto, the ``path'' or ``tour'' version of directed orienteering, where one wants a path/tree of maximum profit subject to the cost limit, is much better understood. There are polynomial time approximation algorithms with guarantees $O(\log n)$~\cite{NagarajanR11,SvenssonTV18} and $O(\log^2k)$~\cite{ChekuriKP12}. However, these results do not imply anything for \dto. Unlike  undirected graphs, in the directed case, we cannot go between trees and tours by doubling edges.  

Our algorithm for \dto in fact follows as a special case of a more general algorithm for  the {\em submodular tree orienteering} (\sto) problem. Here, we are also given a monotone submodular function $f:2^V\rightarrow \R_+$ on the vertex set, and the goal is to find an arborescence containing vertices $T\sse V$ that maximizes $f(T)$ subject to the cost limit.  The ``tour'' or ``path'' version of submodular orienteering was studied previously in \cite{ChekuriP05}, where a quasi-polynomial time $O(\frac{\log k}{\log\log k})$-approximation algorithm was obtained. While we rely on many  ideas from \cite{ChekuriP05}, we also need a number of  new ideas- as discussed next.  

Interestingly, our techniques can be easily extended to obtain tight quasi-polynomial time approximation algorithms for several other directed network design problems such as polymatroid Steiner tree, single-source buy-at-bulk and priority Steiner tree. 

\subsection{Results and Techniques}\label{sec:results}
Our main result is:
\begin{theorem}\label{thm:main}
There is an $O(\frac{\log k}{\log\log k})$-approximation algorithm for submodular tree orienteering that runs in $(n\log B)^{O(\log^{1+\epsilon} k)}$ time for any constant $\epsilon>0$.
\end{theorem}
The high-level approach here is the elegant ``recursive greedy'' algorithm from \cite{ChekuriP05} for the submodular {\em path} orienteering problem, which in turn is similar to the recursion used in Savitch's theorem~\cite{Savitch70}. In order to find an approximately optimal  $s-t$ path with budget $B$, the algorithm in \cite{ChekuriP05}  {\em guesses} the ``middle node'' $v$ on the optimal $s-t$ path as well as the cost $B'$ of the optimal path segment from $s$ to $v$. Then, it  solves two smaller instances recursively and sequentially:
\begin{enumerate}
    \item find an approximately optimal  $s-v$ path $P_{left}$ with budget $B'$.
    \item  find an approximately optimal  $v-t$ path $P_{right}$ with budget $B-B'$ that  {\em augments} $P_{left}$. 
\end{enumerate}
Clearly, the depth of recursion is $\log_2 k$ where $k$ denotes the number of nodes in an optimal path. The key step in the analysis is to show that the approximation ratio is equal to the  depth of recursion. 
In the {\em  tree} version that we consider there are two additional issues:
\begin{itemize}
    \item Firstly, there is no middle node $v$ in an arborescence. A natural choice is to consider  a balanced separator node as $v$: it is well known that any tree has a $\frac13-\frac23$ balanced separator. Indeed, this is what we  use. Although, this leads to  an imbalanced recursion (not exactly half the nodes in each subproblem), the {\em maximum} recursion depth is still $O(\log k)$ and we show that the approximation ratio can be bounded by this quantity. 

    \item Secondly (and more importantly), we cannot simply concatenate the solutions to the two subproblems. If $r$ is the root of the original instance, the two subproblems involve arborescences rooted at $r$ and $v$ respectively. In order to finally obtain an $r$-arborescence, we need to additionally ensure that the  subproblem with root $r$ returns an arborescence containing the separator node $v$, and such requirements can accumulate recursively! Fortunately, there is a clean solution to this issue. We generalize the recursion by also specifying a ``responsibility'' subset $Y\sse V$ for each subproblem, which means that the resulting arborescence {\em must} contain all nodes in $Y$. Crucially, we can show that the size of any responsibility subset is bounded by the       recursion depth $d=O(\log k)$. This allows us to implement the recursive step by additionally {\em guessing} how the 
    responsibility subset $Y$ is passed on to the two subproblems. The number of such guesses is at most $2^d=poly(k)$, and so the overall time remains quasi-polynomial. The responsibility subset $Y$  is empty at the highest level of recursion and  has size at most one at the lowest level of recursion: $|Y|$ may increase and decrease in the intermediate levels.  
\end{itemize}

A direct consequence of Theorem~\ref{thm:main} is an 
 $O(\frac{\log  k}{\log\log k})$-approximation algorithm for \dto and an $O(\frac{\log^2 k}{\log\log k})$-approximation  for directed Steiner tree. This matches the previous best  bound (in quasi-polynomial time) for directed Steiner tree~\cite{GLL18}. However, our approach is much simpler and also achieves a  better exponent in the running time: our time is $n^{O(\log^{1+\epsilon}k)}$ whereas the previous algorithm required $n^{O(\log^{5}k)}$ time. \cite{GLL18} also showed that one cannot obtain an $o(\log^2k/\log\log k)$-approximation ratio for directed Steiner tree in quasi-polynomial time unless $NP\sse \cap_{0<\epsilon<1} ZPTIME(2^{n^\epsilon})$. Hence Theorem~\ref{thm:main} is also tight under the same assumption.  


Another application of Theorem~\ref{thm:main} is to the directed polymatroid Steiner tree problem, where there is a matroid with groundset $V$ (same as the vertices) and one needs to find a minimum cost arborescence that visits some base of the matroid. We obtain an $O(\log^2k/\log\log k)$-approximation, which improves over the previous best $O(\log^3 k)$ bound~\cite{CalinescuZ05}. 

We also extend our main result (Theorem~\ref{thm:main}) to a setting with additional length constraints. In addition to the input to \sto, here we are given a length function $\ell:E\rightarrow \mathbb{Z}_+$ and a bound $L$.   The goal here is to find an arborescence on vertices $T$ maximizing $f(T)$ where (i) the cost of edges in $T$ is at most $B$ and (ii) the sum $\sum_{v\in T} \ell_T(v)\le L$ where $\ell_T(v)$ is the length of the $r-v$ path in $T$. We assume that the lengths are polynomially bounded. Our technique can be extended to:
\begin{theorem}\label{thm-lc}
There is an $O(\frac{\log k}{\log \log k})$-approximation algorithm that runs in quasi-polynomial time for the submodular tree orienteering problem with length constraints.
\end{theorem}
This algorithm follows a similar recursive structure as for \sto, where we guess and maintain some additional quantities: the length budget $L'$ available to the  subproblem and a bound $D(v)$ on the length of the $r-v$ path for each vertex $v$ in the responsibility subset $Y$. This idea can also be used to obtain an 
$O(\frac{\log k}{\log \log k})$-approximation for the variant of \sto with hard deadlines on length (see Section~\ref{subsec:sto-deadline} for details). 

As a direct application of Theorem~\ref{thm-lc}, we obtain an $O(\frac{\log^2 k}{\log\log k})$-approximation  for single-source buy-at-bulk network design. This  improves over  the previous best $O(\log^3k)$-approximation  \cite{Anton11}. Buy-at-bulk network design is a well-studied  generalization of Steiner tree  that involves concave cost-functions on edges. See Section~\ref{subsec:bab} for more details.  Our result holds for the harder ``non uniform'' version of the problem, where cost-functions may differ across edges.

Another application of Theorem~\ref{thm-lc} is to the priority Steiner tree problem, where edges/terminals have priorities (that represent quality-of-service) and the path for each terminal must contain edges of at least its priority. We obtain a quasi-polynomial time $O(\frac{\log^2 k}{\log\log k})$-approximation  even for this problem. We are not aware of any previous result for  directed priority Steiner tree. 

It follows from  the hardness result in \cite{GLL18} that  all our approximation ratios are tight (up to constant factors) 
assuming $NP \not\sse \cap_{0<\epsilon<1} ZPTIME(2^{n^\epsilon})$.

\subsection{Related Work}
The first quasi-polynomial time algorithm for directed Steiner tree was given by \cite{CharikarCCDGGL99}, where an $O(\log^3 k)$-approximation ratio was obtained. This was a recursive algorithm that has a very different structure than ours. The natural cut-covering LP relaxation of directed Steiner tree was shown to have an $\Omega(\sqrt{k})$ integrality gap by \cite{ZosinK02}. Later, \cite{FriggstadKKLST14} showed that one can also obtain an $O(\log^3k)$-approximation ratio relative to the $O(\log k)$-level Sherali-Adams lifting of the natural LP. (Previously, \cite{Rothv11} used the stronger Lasserre hierarchy to obtain the same approximation ratio.) 
Very recently, \cite{GLL18} improved the approximation ratio to $O(\log^2k/\log\log k)$, still in quasi-polynomial time. Their approach was to reduce directed Steiner tree to a new problem, called ``label consistent subtree'' for which they provided an $O(\log^2k/\log\log k)$-approximation algorithm (in quasi-polynomial time) by rounding a Sherali-Adams LP. In contrast, we take a simpler and more direct approach by extending the recursive-greedy algorithm of \cite{ChekuriP05}.

A well-known special case of directed Steiner tree is the group Steiner tree problem~\cite{GargKR00}, for which the best polynomial-time approximation ratio is $O(\log^2k \log n)$. This is relative to the natural LP relaxation. A combinatorial algorithm with slightly worse approximation ratio was given by \cite{ChekuriEK06}. In quasi-polynomial time, there is an $O(\log^2k/\log\log k)$-approximation algorithm, which follows from \cite{ChekuriP05}. There is also an $\Omega(\log^{2-\epsilon}k)$-hardness of approximation for group Steiner tree~\cite{HalperinK03}. Recently, \cite{GLL18} showed that this reduction can be refined to prove an $\Omega(\log^2k/\log\log k)$-hardness of approximation.

\cite{CalinescuZ05} considered a polymatroid generalization of both undirected and directed Steiner tree. For the directed version, they obtained an $O(\log^3k)$-approximation in quasi-polynomial time by extending the approach of \cite{CharikarCCDGGL99}. We improve this ratio to $O(\log^2k/\log\log k)$, which is also the best possible. It is unclear if one can use LP-based methods such as \cite{GLL18} to address this problem. 

Buy-at-bulk network design problems, that involve concave cost-functions,  have been studied extensively as they model economies of scale (which is common in several applications). In the undirected case, a constant-factor approximation algorithm is known for {\em uniform} single-source buy-at-bulk~\cite{GMM09} and an $O(\log k)$-approximation algorithm is known for  the {\em non-uniform} version~\cite{MMP08}. The non-uniform problem is also hard to approximate better than $O(\log\log n)$ ~\cite{CGNS08}. For the directed case that we consider, the only prior result is \cite{Anton11} which implies a quasi-polynomial time $O(\log^3k)$-approximation for the non-uniform version. Buy-at-bulk problems have also been studied for multi-commodity flows~\cite{CHKS10}, which we do not consider in this paper.

The priority Steiner problem was introduced to model quality-of-service requirements in networking~\cite{CNS04}. It is fairly well-understood in the undirected setting: the best approximation ratio known  is $O(\log n)$~\cite{CNS04} and it is  $\Omega(\log\log n)$ hard-to-approximate~\cite{CGNS08}. 



\subsection{Preliminaries}
The input to  the submodular tree orienteering (\sto) problem consists of (i)  a directed graph $G= (V, E)$ with edge costs $c: E \to \mathbb{Z}_+$, (ii) root vertex $r^{*}\in V$, (iii) a budget $B\ge 0$ and (iv)   a monotone submodular function $f: 2^V \to \mathbb{R}_+$ on the power set of the vertices. As usual, we may assume (without loss of generality) that the underlying graph is complete and the costs $c$ satisfy triangle inequality.  We assume throughout that all edge costs are integer valued. We also use the standard value-oracle model for submodular functions, which means that our algorithm can access the value $f(S)$ for any $S \subseteq V$ in constant time. Finally, we assume that for all $S \sse V$, $f(S)$ is polynomially bounded in $n=|V|$. 

The goal in \sto  is to find an out-directed arborescence $T^*$ that is rooted at $r^{*}$ and  maximizes $f(V(T^*))$ such that the cost of edges in $T^*$ is less than $B$, i.e. $\sum_{e \in E(T^*)}c(e) \leq B$. Henceforth, we will use $f(V(T))$ and $f(T)$ interchangeably to mean $f$ evaluated at the vertex set of $T$.

\section{Algorithms for Submodular Tree Orienteering}
We first descibe the basic algorithm that leads to an $(nB)^{O(\log k)}$ time $O(\log k)$-approximation algorithm for \sto in Section~\ref{subsec:main}. This already contains the main ideas. Then, in Section~\ref{subsec:qp} we show how to make the algorithm truly quasi-polynomial time by implementing it in $(n \log B)^{O(\log k)}$ time. Finally, in Section~\ref{subsec:imp} we show how to obtain a slightly better $O(\frac{\log k}{\log\log k})$ approximation ratio in  $(n \log B)^{O(\log^{1+\epsilon} k)}$ time. 

\subsection{The Main Algorithm}\label{subsec:main}
The procedure $\text{RG}(r, Y, B, X, i)$ implements the algorithm.
\begin{itemize}
\item The parameters $r\in V$ and $B\ge 0$ denote that we are searching for an  $r$-rooted arborescence with  cost at most $B$.
\item $Y\sse V$ is a set of vertices that {\em must} be visited from $r$. We refer to set  $Y$ as the  responsibilities for this subproblem.
\item The parameter $X\sse V$ indicates that we aim to  maximize function $f_X(T) = f(T \cup X) - f(X)$; that is we seek to find an arborescence that {\em augments} a given  set $X$.
\item The parameter $i\ge 1$ indicates the depth of recursion allowed and that the arborescence   returned can contain at most $(\frac{3}{2})^i$ vertices, excluding the root.
\end{itemize}

\begin{algorithm}[h]
\caption{RG$(r, Y, B, X, i)$}\label{rec-greedy}
\begin{algorithmic}[1]
\If {$(|Y| > (\frac{3}{2})^i)$} \Return {\bf Infeasible} \EndIf
\If {$i = 1$} 
\If {($|Y| = 0$)}  \Comment{No responsibility for $r$}
\State pick $v \in V: c(r, v) \leq B$   that maximizes $f_X(v)$ \Comment{Guess base-case vertex} \EndIf
\If {($|Y|$ = 1)}  \Comment{$r$ must visit vertex $v \in Y$}
\If {$(c(r, v) \leq B)$} \Return $\{(r, v)\}$ 
\Else \ \Return  {\bf Infeasible}
\EndIf
\EndIf
\EndIf
\State $T \leftarrow \emptyset$
\State $m \leftarrow f_X(\emptyset)$
\For{each $v \in V$} \Comment{Guess separator vertex}
\For{$S \subseteq Y$} \Comment{Guess responsibilities for left/right subtrees}
\For{$1 \leq B_1 \leq B$} \Comment{Guess subtree budget}
\State $T_1 \leftarrow \text{RG}(r, (S \cup \{v\}) \setminus \{r\}, B_1, X, i-1)$
\State $T_2 \leftarrow \text{RG}(v, Y \setminus (S \cup \{v\}), B - B_1, X \cup V(T_1), i-1)$
\If {$(f_X(T_1 \cup T_2) > m)$} 
\State $T \leftarrow T_1 \cup T_2$
\State $m \leftarrow f_X(T)$
\EndIf
\EndFor
\EndFor
\EndFor
\State \Return $T$
\end{algorithmic}
\end{algorithm}

\begin{remark} \label{rem-alg1}
Given a valid input to the \sto problem, our solution is $T \leftarrow \text{RG}(r^*, \emptyset, B, \emptyset, d)$ for $d \geq \log_{3/2}k$ where $k$ is the number of vertices in an optimal solution.
\end{remark}

\begin{fact} \label{fact-vertex-sep} 
Any tree on $n$ vertices has a vertex $v$ whose removal leads to each  connected components  having size at most $n/2$. These  components can be clubbed together to form two  connected components (both containing $v$), each of  size  at most $2n/3$.
\end{fact}

\begin{proposition}
The maximum size of set $Y$ in any subproblem of $\text{RG}(r, \emptyset, B, \emptyset, d)$ is $d$.
\end{proposition}
\textit{Proof.} To prove the above statement, we argue that the invariant $|Y| + i \leq d$ holds in every subproblem of $\text{RG}(r, \emptyset, B, \emptyset, d)$ of the form $\text{RG}(r, Y, B, X, i)$. We prove this by induction on $i$. For the base case, let $i = d$. In this case, $|Y| = 0$, and thus the aforementioned invariant clearly holds. Inductively, assume that the invariant holds at some depth $i > 1$ for some responsibility set $Y$. Let $\text{RG}(r', Y', B', X', i-1)$ be a subproblem of $\text{RG}(r, Y, B, X, i)$. From the description of the algorithm, we can see that the size of $Y$ increases by at most $1$ in any subproblem: so $|Y'|\le 1+|Y|$. Combining this observation with the induction hypothesis $|Y| + i \leq d$, we get $|Y'| + i-1 \leq |Y| + 1 + i-1 \leq d$ which completes the induction. 

Finally, as $i \geq 1$, we have $|Y| < d$ in any subproblem of $\text{RG}(r, \emptyset, B, \emptyset, d)$. \hspace*{\fill} \qedsymbol


\begin{proposition} \label{prop-alg1}
The running time of the procedure $\text{RG}(r, Y, B, X, i)$ is ${O}((nB \cdot 2^{d+2})^{i})$.
\end{proposition}
\textit{Proof.} We prove the above claim by induction on $i$. Let us denote the running time of $\text{RG}(r, Y, B, X, i)$ by $T(i)$. We want to show that $T(i) \le c\cdot (nB \cdot 2^{d+2})^{i}$ for some fixed constant $c$.  For the base case, let $i = 1$. From the description of the procedure, we can see that when $i = 1$, it only performs a linear number of operations. Thus $T(1) = {O}(n)$ which proves the base case. Inductively, assume that the claim holds for all values $i' < i$. From the description of the procedure, we have the following recurrence relation: $T(i) = nB\cdot 2^d (2T(i-1) + {O}(n))$. This follows from the fact that we have $n$ guesses for the separator vertex, $B$ guesses for the split in the cost of the left and right subtree and at most $2^d$ guesses on the responsibility set assigned to each subtree (since $|Y| \leq d$). For every combination of the guesses, we make $2$ recursive calls. Applying the induction hypothesis, we get $T(i) = nB\cdot 2^d (2 \cdot  (c\cdot (nB \cdot 2^{d+2})^{i-1}) + {O}(n))\le c\cdot (nB \cdot 2^{d+2})^{i}$ which completes the induction. \hspace*{\fill} \qedsymbol

\begin{lemma} \label{lem-alg1}
Let $T$ be the arborescence returned by $\text{RG}(r, Y, B, X, i)$. Let $T^*$ be a compatible arborescence for the parameters $(r, Y, B, X, i)$, i.e. $T^*$ is an $r$-rooted arborescence that visits all vertices in $Y$, and contains at most $(\frac{3}{2})^i$ non-root vertices with a total cost of at most $B$ . Then $f_X(T) \geq f_X(T^*)/i$.
\end{lemma}
\textit{Proof.} We prove the lemma by induction on $i$. For the base case, let $i=1$. Since $T^*$ is feasible for $i=1$,  $T^*$ is either empty or contains a single edge. If $|Y| = 0$, then we guess the base-case vertex and return the one that maximizes $f_X$ subject to the given budget: so $f_X(T) \geq f_X(T^*)$ in this case. If $|Y| = 1$, then $T^*$ has a single edge, say $(r, v)$. Our procedure here will return the  arborescence with $(r, v)$, and so $f_X(T) = f_X(T^*)$. Thus, in either case, we get $f_X(T) \geq f_X(T^*)$ which proves the base case.

Suppose that $i > 1$. Let $v$ be the  vertex in $T^*$ obtained from Fact~\ref{fact-vertex-sep} such that we can separate $T^*$ into two connected components: $T_1^*$ containing $r$ and $T_2^*= T^* \setminus T_1^*$, where  $\max(|V(T_1^*)|, |V(T_2^*)|) \leq \frac{2}{3}|V(T^*)|$. Note that $T^*_1$ is an $r$-rooted arborescence that contains $v$ and $T^*_2$ is a $v$-rooted arborescence. 
Let $Y_2 \subseteq Y \setminus \{v\}$ be those vertices of $Y\setminus v$ that are contained in $T_2^*$, and let  $Y_1 = Y \setminus Y_2$. Because $T^*$ contains $Y$, it is clear that $\{v\}\cup Y_1\cup Y_2\supseteq Y$.  Finally, 
let $c(T_1^*) = B_1$ and $c(T_2^*) = B_2 \leq B - B_1$. Note also that  
 $|V(T^*) \setminus \{r\}| \leq (\frac{3}{2})^i$. By the property of the separator vertex $v$, $\max(|V(T_1^*)|, |V(T_2^*)|) \leq \frac{2}{3}|V(T^*)| \leq (\frac{3}{2})^{i-1} + \frac{2}{3}$. Excluding the root vertex in $T_1^*$ and $T_2^*$, the number of non-root vertices in either arborescence is $\leq (\frac{3}{2})^{i-1}$. We can thus claim that:
 \begin{equation}\label{eq:left-rec}
 T_1^* \mbox{ is compatible with }(r, Y_1 \cup \{v\} \setminus \{r\}, B_1, X, i-1) \mbox{ and}     
 \end{equation}
 \begin{equation}\label{eq:right-rec}
   T_2^*   \mbox{ is compatible with } (v, Y \setminus (Y_1 \cup \{v\}), B - B_1, X \cup V(T_1), i-1).
 \end{equation}
Now consider the call $\text{RG}(r, Y, B, X, i)$. 
Since we iteratively set every vertex to be the separator vertex, one of the guesses is $v$. Moreover, we iterate over all subsets $S \subseteq Y$, and thus some guess must set $S = Y_1$. Since $B_1 \leq B$, we also correctly guess $B_1$ in some iteration. Thus, we see that one of the set of calls made is 
$$T_1 \leftarrow \text{RG}(r, Y_1 \cup \{v\} \setminus \{r\}, B_1, X, i-1) \mbox{ and }T_2 \leftarrow \text{RG}(v, Y \setminus (Y_1 \cup \{v\}), B - B_1, X \cup V(T_1), i-1)$$

We now argue that $T = T_1 \cup T_2$ has the property that $f_X(T) \geq f_X(T^*)/i$. By \eqref{eq:left-rec} and  induction,  
\begin{equation} \label{eq-t1}
f_X(T_1) \geq \frac{1}{i-1}f_X{(T_1^*)}
\end{equation}

Let $X' = X \cup V(T_1)$. Similarly, by \eqref{eq:right-rec} and  induction, we have
\begin{equation} \label{eq-t2}
f_{X'}(T_2) \geq \frac{1}{i-1}f_{X'}{(T_2^*)}
\end{equation} 

The rest of this proof is identical to a corresponding result in \cite{ChekuriP05}. We have $f_{X'}(T_2^*) = f(T_2^* \cup T_1 \cup X) - f(T_1 \cup X) = f_X(T_1 \cup T_2^*) - f_X(T_1)$. Using this in \eqref{eq-t2}, we get 
\begin{align}
f_{X'}(T_2) & \geq \frac{1}{i-1}(f_X(T_1 \cup T_2^*) - f_X(T_1)) \notag \\
		 & \geq \frac{1}{i-1}(f_X(T_2^*) - f_X(T)) \label{eq-t2-2}
\end{align}
where the last inequality follows from the monotonicity of the function $f$.

We see that $f_X(T) = f_X(T_1 \cup T_2) = f(T_1 \cup T_2 \cup X) - f(X) + f(T_1 \cup X) - f(T_1 \cup X) = f_X(T_1) + f_{X'}(T_2)$. Thus using \eqref{eq-t1} and \eqref{eq-t2-2}, we get 
$$f_{X}(T)\quad \geq \quad \frac{1}{i-1}(f_X(T_1^*) + f_X(T_2^*) - f_X(T)) \quad \geq\quad  \frac{1}{i-1}(f_X(T^*) - f_X(T)) $$
where the last inequality follows by the submodularity of $f$. On rearranging the terms, we get 
\begin{equation}
f_X(T) \geq \frac{1}{i}f_X(T^*) \notag
\end{equation}
which concludes the induction. \hspace*{\fill} \qedsymbol

Remark \ref{rem-alg1}, Proposition \ref{prop-alg1} and Lemma \ref{lem-alg1} imply:
\begin{theorem}
There is a  $(\log_{1.5} k)$-approximation algorithm for the submodular tree orienteering problem that runs in time ${O}(nB)^{{O}(\log k)}$.
\end{theorem}


\subsection{Quasi-Polynomial Time Algorithm}\label{subsec:qp}
Here we show how our algorithm can be implemented more efficiently in $(n\log B)^{O(\log k)}$ time. The idea here is the same as  \cite{ChekuriP05}, but applied on top of Algorithm~\ref{rec-greedy}. 

\begin{algorithm}[h]
\caption{RG-QP$(r, Y, B, X, i)$}\label{rec-greedy-qp}
\begin{algorithmic}[1]
\If {$(|Y| > (\frac{3}{2})^i)$} \Return {\bf Infeasible} \EndIf
\If {($i = 1$)} 
\If {($|Y| = 0$)}  \Comment{No responsibility for $r$}
\State pick $v \in V: c(r, v) \leq B$   that maximizes $f_X(v)$ \Comment{Guess base-case vertex}
\EndIf
\If {($|Y|$ = 1)}  \Comment{$r$ must visit vertex $v \in Y$}
\If {$(c(r, v) \leq B)$} \Return $\{(r, v)\}$ 
\Else \ \Return  {\bf Infeasible}
\EndIf
\EndIf
\EndIf
\State $T \leftarrow \phi$
\State $m \leftarrow f_X(\phi)$
\For{each $v \in V$} \Comment{Guess separator vertex}
\For{$S \subseteq Y$} \Comment{Guess responsibilities for left/right subtrees}
\For{$1 \leq u \leq U$} \Comment{Guess subtree function value}
\State $B_1 \leftarrow \min_b(\text{RG-QP}(r, (S \cup \{v\}) \setminus \{r\}, b, X, i-1) \geq u)$ \Comment{Binary search for  $B_1$}
\If{($B_1 = \infty$)} continue \EndIf
\State $T_1 \leftarrow \text{RG-QP}(r, (S \cup \{v\}) \setminus \{r\}, B_1, X, i-1)$
\State $T_2 \leftarrow \text{RG-QP}(v, Y \setminus (S \cup \{v\}), B - B_1, X \cup V(T_1), i-1)$
\If {$(f_X(T_1 \cup T_2) > m)$} 
\State $T \leftarrow T_1 \cup T_2$
\State $m \leftarrow f_X(T)$
\EndIf
\EndFor
\EndFor
\EndFor
\State \Return $T$
\end{algorithmic}
\end{algorithm}

The key idea here is that we no longer iterate through all values in $[1, B]$ to guess the recursive budget $B_1$. Instead, the step $B_1 \leftarrow \min_b(\text{RG-QP}(r, (S \cup \{v\}) \setminus \{r\}, b, X, i-1) \geq u)$ is implemented as a binary search over the range $[1, B]$. Here we assume that $U$ is an upper bound on the function value. The following results are straightforward extensions of those in Section~\ref{subsec:main}.

\begin{proposition} \label{prop-alg2}
The running time of the procedure $\text{RG-QP}(r, Y, B, X, i)$ is ${O}((nU \cdot 2^d \cdot \log B)^{i})$.
\end{proposition}

\begin{lemma} \label{lem-alg2}
Let $T$ be the arborescence returned by $\text{RG-QP}(r, Y, B, X, i)$. Let $T^*$ be a compatible arborescence for the parameters $(r, Y, B, X, i)$, and $f_X(T^*) \leq U$. Then $f_X(T) \geq f_X(T^*)/i$.
\end{lemma}
The proof of this lemma is similar to Lemma~\ref{lem-alg1} and  can be found in the appendix.  
Combining Proposition \ref{prop-alg2} and Lemma \ref{lem-alg2} and using polynomially bounded profits, we obtain: 
\begin{theorem}
There is an ${O}(\log k)$-approximation algorithm for the submodular tree orienteering problem that runs in time ${O}(n \log B)^{{O}(\log k)}$. 
\end{theorem}

\subsection{Improved Approximation Ratio}
\label{subsec:imp}
Here we show how to reduce the depth of our recursion at the cost of additional guessing. The high-level idea is the same as a similar result in \cite{ChekuriP05}, but we need some more care because our recursion is more complex. 

Let $s=\epsilon\cdot \log\log k$ where $\epsilon>0$ is some fixed constant. At each level of recursion, our new algorithm will guess all relevant quantities in $s$ levels of the recursion in Algorithm~\ref{rec-greedy}. So the new recursion depth will be $d/s = O(\frac{\log k}{\log\log k})$ where $d=O(\log k)$ was the old depth. 
Recall that the number of subproblems generated at each level of recursion in Algorithm~\ref{rec-greedy} is $2n2^d B$. Since we want to generate all subproblems in the next $s$ levels, each subproblem in the new algorithm generates $(2n2^d B)^{2^s}$ many   subproblems. As $d=O(\log k)$ and $s=\epsilon \cdot  \log\log k $, the overall  running time for the new  algorithm is at most $(2n2^d B)^{2^s\, d}\le (nB)^{O(\log^{1+\epsilon} k)}$.

Next, we will prove a lemma bounding the objective value at each level of the recursion. Below, $i\in \{1,2,\cdots d/s\}$ denotes the depth allowed in any subproblem of the new recursion.

\begin{lemma} \label{lem-alg3}
Let $T$ be the arborescence returned by the improved approximation algorithm for parameters $(r, Y, B, X, i)$. Let $T^*$ be some arborescence compatible with the same parameters. Then $f_X(T) \geq f_X(T^*)/i$.
\end{lemma}
\textit{Proof. }We will prove the claim by induction on $i$. For the base case, let $i =1$. This is equivalent to the $s^{th}$ level of the earlier algorithms, which implies that $T^*$ contains at most $(\frac{3}{2})^s$ vertices excluding the root vertex. Since we guess all parameters for $s$ levels of recursion, there exist guesses such that we can write $T^* = \cup_{j=0}^{2^s} T_j^*$ such that each $T_j^*$ is either empty or contains a single edge. Since the edges in $T_j^*$ are compatible with our guesses, and we will pick the best possible edge for $T_j$, we can conclude that $f_X(T) \geq f_X(T^*)$ which proves the base case.

Fix some $i > 1$. Consider the call to the new algorithm with the parameters $(r, Y, B, X, i)$ where $i$ denotes the new depth. Since $T^*$ is compatible with the given parameters, one can iteratively obtain a choice of  separator node $v$, responsibility set $S$ and budget $B'$ at each subproblem in the next $s$ levels (exactly as in Lemma~\ref{lem-alg1}). This allows us to  write $T^* = \cup_{j=1}^{2^s} T_j^*$ such that each $T_j^*$ is compatible with some subproblem at new depth $(i-1)$. 
For each $j=1,\cdots 2^s$ let $T_j$ denote the solution returned by the $j^{th}$ subproblem. The solution to the current subproblem is then $T=\cup_{j=1}^{2^s} T_j$. By induction, we have that $f_X(T_j) \geq f_{X_j}(T_j^*)/(i-1)$ where $X_j = X \cup (\bigcup_{a=1}^{j-1}T_a)$. 
Let $h = 2^s$.  We will show below that
\begin{equation}
    \label{eq:rec-multi}
\sum_{j=1}^{h} f_{X_j}(T^*_j) \geq f_X(T^*) - f_X(T). 
\end{equation}
This would imply
 $$f_X(T)  = \sum_{j=1}^h f_{X_j}(T_j)  \geq \sum_{j=1}^h \frac{f_{X_j}(T^*_j)}{(i-1)} \geq \frac{1}{(i-1)} (f_X(T^*) - f_X(T)),$$ which upon  rearranging terms yields $f_X(T)\ge f_X(T^*)/i$ as  desired.
 
 To prove \eqref{eq:rec-multi} consider 
\begin{align*}
\sum_{j=1}^{h} f_{X_j}(T_j^*) + f_X(T) &= \sum_{j=1}^{h-1} f_{X_j}(T_j^*) + f_{X_h}(T_h^*) + f_X(T) \\
							  &= \Big(\sum_{j=1}^{h-1} f_{X_j}(T_j^*)\Big) + f(T_h^* \cup X \cup (\bigcup_{j=0}^{h-1}T_j)) - f(X \cup (\bigcup_{j=0}^{h-1}T_j)) + f(T \cup X) - f(X) \\
							  &\text{ applying submodularity to the $2^{nd}$ and $4^{th}$ term} \\
							  &\geq \Big(\sum_{j=1}^{h-1} f_{X_j}(T_j^*)\Big) + f(T_h^* \cup X \cup T) + f(X \cup (\bigcup_{j=0}^{h-1}T_j)) - f(X \cup (\bigcup_{j=0}^{h-1}T_j)) - f(X) \\
							  &= \Big(\sum_{j=1}^{h-1} f_{X_j}(T_j^*)\Big) + f(T_h^* \cup X \cup T) - f(X) \\
							  &\text{ inductively for all $k=h-1,\cdots 1,0$ using the same steps as above} \\
							   &  \geq \Big(\sum_{j=1}^{k} f_{X_j}(T_j^*)\Big) + f((\bigcup_{j=k+1}^h T_j^*) \cup X \cup T) - f(X) \\ 
							  &\geq f(T^* \cup T \cup X) - f(X)\\
							  &\text{ and from the monotonicity of $f$ this is } \\	
							  &\geq f(T^* \cup X) - f(X)	= f_X(T^*). 
\end{align*}
This completes the proof. \hspace*{\fill} \qedsymbol


We further improve the runtime by applying the binary-search idea described in Section \ref{subsec:qp}. Combining this with Lemma \ref{lem-alg3} and using polynomially bounded profits, we obtain Theorem~\ref{thm:main}. 

\section{Applications} \label{sec:app}
\paragraph{Directed tree orienteering (\dto)} This is the special case of \sto when the reward function is linear, i.e. of the form $f(S)=\sum_{v\in S} p_v$ where each $v\in V$ has reward $p_v\in \mathbb{Z}$. So Theorem~\ref{thm:main} applies directly to yield a quasi-polynomial time ${O}(\frac{\log k}{\log \log k})$-approximation algorithm. To the best of our knowledge,  no non-trivial approximation ratio followed from prior techniques.  

\paragraph{Directed Steiner tree} Here, we are given a graph $(V,E)$ with edge costs $c\in \R^E_+$, root $r$ and a subset $U\sse V$ of terminals. The goal is to find an $r$-rooted arborescence that contains all of $U$ and minimizes the total cost. By shortcutting over non-terminal vertices of degree at most two, we can assume that there is an optimal solution where every non-terminal vertex has degree at least three. So there is an optimal solution containing at most $2k$ vertices where $k=|U|$ is the number of terminals. We can use a standard set-covering approach to solve directed Steiner tree using \dto. We first guess (up to factor 2) a bound $B$ on the optimal cost. Then we iteratively run the \dto algorithm with budget $B$ and reward of one for all {\em uncovered} terminal vertices. Assuming that the bound $B$ is a correct guess, the optimal value of each \dto instance solved above equals $k'$, the number of uncovered terminals. As we use a $\rho={O}(\frac{\log k}{\log \log k})$ approximation for \dto, the number of iterations before covering all terminals is at most $O(\rho\cdot \log k)$. Using Theorem~\ref{thm:main}, this implies:
\begin{theorem}\label{thm:steiner}
There is a deterministic ${O}(\frac{\log^2 k}{\log \log k})$-approximation algorithm for directed Steiner tree in $n^{O(\log^{1+\epsilon} k)}$ time, for any constant $\epsilon>0$.
\end{theorem}
Our approximation ratio matches that obtained recently~\cite{GLL18}. Our algorithm is deterministic and has a better running time:  the algorithm in \cite{GLL18} requires $n^{O(\log^5 k)}$ time. Moreover, our approach is much simpler. However, we note that an LP relaxation based approach as in \cite{GLL18} may have  other advantages. 

\paragraph{Polymatroid Directed Steiner tree} This problem was introduced in \cite{CalinescuZ05} with applications in sensor networks. As before, we are given a directed graph $(V,E)$ with edge costs $c\in \R^E_+$ and root $r$. In addition, there is a matroid defined on groundset $V$ (same as the vertices) and the goal is to find a min-cost arborescence rooted at $r$ that contains some base of the matroid. As matroid rank functions are submodular (and integer valued), we can apply Theorem~\ref{thm:main} to obtain an 
${O}(\frac{\log k}{\log \log k})$-approximation algorithm for the corresponding \sto instance (reward-maximization), where $k\le |V|$ is the rank of the matroid. We then use  a set-covering approach as outlined above, that  iteratively solves   \sto instances until the set of covered vertices contains a base of the matroid. Crucially, the contraction of any matroid is another matroid: so the function $f$ used in each such \sto instance is still a matroid rank function. This yields  an ${O}(\frac{\log^2 k}{\log \log k})$-approximation algorithm for polymatroid Steiner tree as well. 
This result improves over the $O(\log^3k)$ ratio in \cite{CalinescuZ05}. 

\section{Extensions of  Submodular Tree Orienteering}\label{sec:extend}

In this section, we will consider two extensions of \sto that involve  additional length constraints. We then use this extension to obtain an improved approximation algorithm for directed buy-at-bulk network design and priority Steiner tree. 
 
\subsection{\sto with Length Constraints}\label{subsec:sto-lc}
For the first extension, along with the input to \sto, we are given a length function $\ell: E \to \mathbb{Z}_{+}$, and an additional bound $L$. 
Note that in an arborescence, given a vertex $v$, there is a unique path from the root to $v$. Let $p_T(v)$ denote the path from the root $r^*$  to vertex $v$ in arborescence $T$, and   $l_T(v) = \sum_{e \in p_T(v)} \ell(e)$ represents the length of this path. The length constraint requires  the sum of path  lengths $l_T(v)$ to be at most $L$. More formally, the goal now is to find an out-directed arborescence $T^*$ rooted at $r^{*}$ maximizing $f(T^*)$ such that $c(T^*) \leq B$ and $\sum_v l_{T^*}(v) \leq L$. We will refer to this problem as \sto with length constraints.

The main algorithm is implemented by the procedure RG-DC$(r, k_r, Y, D, B, L, X, i)$. The parameters $r, Y, B, X, i$ are the same as described earlier. Recall that in the case of \sto, the idea was to guess a separator vertex $v$, and to guess the bound $B_1$ and $B-B_1$ for the two arborescences that are rooted at $r$ and $v$ respectively, and to assign responsibilities for each arborescence. For \sto with length constraints, we will additionally guess the bound $L_1$ and $L-L_1$ on the sum of lengths for these arborescences. Since the lengths also involve information from the computed subproblems, we add a parameter $k_r$ in the recursive call which denotes the length from $r^{*}$ (the original root) to $r$ (root of the current subproblem). Another subtlety is that even though we guess the bounds $L_1$ and $L - L_1$ correctly, the length from $r$ to $v$ obtained by the procedure may not be the same as in an optimal solution. This is crucial since the length to $v$ will affect the bound of the arborescence rooted at $v$. To get around this, every time we guess a separator vertex $v$, we additionally guess a length bound to get from $r$ to $v$, and add this guess to a dictionary $D$. An arborescence is feasible only if it respects the guessed lengths in $D$.

The procedure RG-DC$(r, k_r, Y, D, B, L, X, i)$ implements the algorithm. First, we provide a description of the parameters.
\begin{itemize}
\item The parameters $r\in V$ and $B\ge 0$ denote that we are searching for an $r$-rooted arborescence with cost at most $B$.
\item The parameter $k_r$ indicates that the length of the path from $r^{*}$ to $r$ is $k_r$. Combined with this $k_r$, we want the sum of lengths in the $r$-rooted arborescence to be at most the bound $L$. Formally, for any feasible $r$-rooted arborescence $T$, we need $\sum_{v \in T}(k_r + l_{T}(v)) \leq L$.
\item $Y\sse V$ is a set of vertices that {\em must} be visited from $r$. We refer to set $Y$ as the  responsibilities for this subproblem.
\item $D$ is a dictionary of vertex-length pairs for vertices in the responsibility set $Y$. It contains pairs $(w, D(w))$ for all $w \in Y$. The length from $r$ to any $w\in Y$ in the returned arborescence must be at most $D(w)$.
\item The parameter $X\sse V$ indicates that we aim to  maximize function $f_X(T) = f(T \cup X) - f(X)$; that is we seek to find an arborescence that {\em augments} a given  set $X$.
\item The parameter $i\ge 1$ indicates the depth of recursion allowed and that the arborescence returned can contain at most $(\frac{3}{2})^i$ vertices, excluding the root.
\end{itemize}

\begin{algorithm}[h]
\caption{RG-DC$(r, k_r, Y, D, B, L, X, i)$}\label{rec-greedy-dc}
\begin{algorithmic}[1]
\If {$(|Y| > (\frac{3}{2})^i)$} \Return {\bf Infeasible} \EndIf
\If {$i = 1$} 
\If {($|Y| = 0$)}  \Comment{No responsibility for $r$}
\State pick $v \in V: c(r, v) \leq B$ and $k_r + \ell(r, v) \leq L$  that maximizes $f_X(v)$ 
 \EndIf
\If {($|Y|$ = 1)}  \Comment{$r$ must visit vertex $v \in Y$}
\If {$(c(r, v) \leq B$, $k_r + \ell(r, v) \leq L$ and $\ell(r, v) \leq D(v)$)} 
\State \Return $\{(r, v)\}$ 
\Else \ \Return  {\bf Infeasible}
\EndIf
\EndIf
\EndIf
\State $T \leftarrow \emptyset$
\State $m \leftarrow f_X(\emptyset)$
\For{each $v \in V$} \Comment{Guess separator vertex}
\For{$S \subseteq Y$} \Comment{Guess responsibilities for  subtrees}
\For{$1 \leq B_1 \leq B$} \Comment{Guess subtree cost budget}
\For{$1 \leq L_1 \leq L$} \Comment{Guess subtree length budget}
\For{$1 \leq d_1 \leq L$} \Comment{Guess length from $r$ to $v$}
\State $D_1 \leftarrow \{ (w, D(w)): w \in S \}$  \Comment{Length guesses for new responsibility set}
\State $D_1(v) \leftarrow \min\{d_1, D_1(v)\}$ \Comment{Update/add length guess for $v$}
\State $D_2 \leftarrow \{ (w, D(w) - d_1): w \in Y \setminus (S \cup v)  \}$   \Comment{Length guesses for $v$-subtree}
\State $T_1 \leftarrow \text{RG-DC}(r, k_r, (S \cup v) \setminus r, D_1, B_1, L_1, X, i-1)$
\State $T_2 \leftarrow \text{RG-DC}(v, k_r + l_{T_1}(v), Y \setminus (S \cup v), D_2, B - B_1, L - L_1, X \cup V(T_1), i-1)$
\If {$(f_X(T_1 \cup T_2) > m)$} 
\State $T \leftarrow T_1 \cup T_2$
\State $m \leftarrow f_X(T)$
\EndIf
\EndFor
\EndFor
\EndFor
\EndFor
\EndFor
\State \Return $T$
\end{algorithmic}
\end{algorithm}

\begin{proposition}\label{prop-alg3}
The running time of RG-DC$(r, k_r, Y, D, B, L, X, i)$ is $O((nBL^2 \cdot 2^d)^i)$ where $d \geq \log_{3/2} k $.
\end{proposition}
The proof of this fact follows from the analysis of Proposition \ref{prop-alg1}, and hence we will omit it here. 

\begin{lemma}\label{lem-alg3}
Let $T$ be the arborescence returned by RG-DC$(r, k_r, Y, D, B, L, X, i)$. Let $T^*$ be a compatible arborescence for the parameters RG-DC$(r, k_r, Y, D, B, L, X, i)$, i.e. $T^*$ is an $r$-rooted arborescence that visits all vertices in $Y$ while respecting the guessed lengths in $D$, and contains at most $(\frac{3}{2})^i$ non-root vertices. The total cost of $T^*$ is at most $B$, and $\sum_{v \in T^*} (k_r + l_{T^*}(v)) \leq L$. Then $f_X(T) \geq f_X(T^*)/i$.
\end{lemma}
\textit{Proof.} We prove the lemma by induction on $i$. For the base case, let $i=1$. Since $T^*$ is feasible for $i=1$, then $T^*$ is either empty or contains a single edge. If $|Y| = 0$, then we guess the base-case vertex and return the one that maximizes $f$ subject to the given budget and length constraints: so $f_X(T) \geq f_X(T^*)$ in this case. If $|Y| = 1$, then $T^*$ has a single edge, say $(r, v)$. Our procedure here will return the  arborescence with $(r, v)$, and so $f_X(T) = f_X(T^*)$. Thus, in either case, we get $f_X(T) \geq f_X(T^*)$ which proves the base case. 

Suppose that $i > 1$. Let $v$ be the  vertex in $T^*$ obtained from Fact~\ref{fact-vertex-sep} such that we can separate $T^*$ into two connected components: $T_1^*$ containing $r$ and $T_2^*= T^* \setminus T_1^*$, where  $\max(|V(T_1^*)|, |V(T_2^*)|) \leq \frac{2}{3}|V(T^*)|$. Note that $T^*_1$ is an $r$-rooted arborescence that contains $v$ and $T^*_2$ is a $v$-rooted arborescence. Let $Y_2 \subseteq Y \setminus \{v\}$ be those vertices of $Y\setminus v$ that are contained in $T_2^*$, and let  $Y_1 = Y \setminus Y_2$. Since $T^*$ contains $Y$, it is clear that $\{v\}\cup Y_1\cup Y_2\supseteq Y$.  Let $c(T_1^*) = B_1$ and $c(T_2^*) = B_2 \leq B - B_1$. Let $L_1 = \sum_{v \in T_1^*} (k_r + l_{T_1^*}(v))$ be the sum of lengths of the vertices in $T_1^*$, and $L_2$, defined analogously, is the sum of lengths of the vertices in $T_2^*$. Observe that $L_2 \leq L - L_1$, and that all lengths of the vertices in $T_2^*$ will share $l_{T_1^*}(v)$. Define $D_1 = \{ (w, D(w)) : w \in Y_1 \}$ and $D_1(v) = l_{T_1^*}(v)$ (if $D(v) > l_{T_1^*}(v)$, then $T^*$ would not be compatible with the given parameters). Let $D_2 = \{ (w, D(w)-l_{T_1^*}(v)) : w \in Y_2 \}$. Note also that $|V(T^*) \setminus \{r\}| \leq (\frac{3}{2})^i$. By the property of the separator vertex $v$, $\max(|V(T_1^*)|, |V(T_2^*)|) \leq \frac{2}{3}|V(T^*)| \leq (\frac{3}{2})^{i-1} + \frac{2}{3}$. Excluding the root vertex in $T_1^*$ and $T_2^*$, the number of non-root vertices in either arborescence is $\leq (\frac{3}{2})^{i-1}$. Thus  
\begin{equation}
    \label{eq:len-left-rec}
T_1^* \mbox{ is compatible with }(r, k_r, (Y_1 \cup v) \setminus r, D_1, B_1, L_1, X, i-1) \mbox{ and }    
\end{equation}
\begin{equation}
    \label{eq:len-right-rec}
T_2^* \mbox{ is compatible with } (v, Y \setminus (Y_1 \cup v), k_r + l_{T_1^*}(v), D_2, B - B_1, L - L_1, X \cup V(T_1), i-1).
\end{equation}
Now consider the call RG-DC$(r, k_r, Y, D, B, L, X, i)$. Since we iteratively set every vertex to be the separator vertex, one of the guesses is $v$. Moreover, we iterate over all subsets $S \subseteq Y$, and thus some guess must set $S = Y_1$. Since $B_1 \leq B$ and $L_1 \leq L$, we also correctly guess $B_1$ and $L_1$. We iteratively guess $d_1 \leq L$, and thus one of the guesses must be $l_{T_1^*}(v)$. Thus, we see that one of the set of calls made is 
$$T_1 \leftarrow \text{RG-DC}(r, k_r, (Y_1 \cup v) \setminus r, D_1, B_1, L_1, X, i-1) \mbox{  and }$$
$$T_2 \leftarrow \text{RG-DC}(v, k_r + l_v, Y \setminus (Y_1 \cup \{v\}), D_2', B - B_1, L - L_1 X \cup V(T_1), i-1),$$
 where $l_v \leq l_{T_1^*}(v)$, and $D_2' \geq D_2$ (component-wise).

We now argue that $T = T_1 \cup T_2$ has the property that $f_X(T) \geq f_X(T^*)/i$. By \eqref{eq:len-left-rec} and  induction,  
\begin{equation} \label{eq-4t1}
f_X(T_1) \geq \frac{1}{i-1}f_X{(T_1^*)}
\end{equation}

Let $X' = X \cup V(T_1)$. By \eqref{eq:len-right-rec} it follows that   $T_2^*$ is also compatible with  
$$(v, k_r + l_v, Y \setminus (Y_1 \cup \{v\}), D_2', B - B_1, L - L_1, X \cup V(T_1), i-1).$$ Again, by induction we have
\begin{equation} \label{eq-4t2}
f_{X'}(T_2) \geq \frac{1}{i-1}f_{X'}{(T_2^*)}
\end{equation} 
The rest of this proof is identical to the proof of Lemma \ref{lem-alg1}.  \hspace*{\fill} \qedsymbol

Combining Proposition \ref{prop-alg3} and Lemma \ref{lem-alg3} gives us the following.
\begin{theorem}
There is an $O(\log k)$-approximation algorithm for the submodular tree orienteering problem with length constraints that runs in time $O(nBL^2)^{O(\log k)}$.
\end{theorem}

\subsubsection{Improved Runtime and Approximation Guarantee} 
As mentioned in Section \ref{subsec:qp}, since $f(\cdot)$ is assumed to be polynomially bounded in $n$, we can guess an upper bound $U$ on the maximum function value. We can then guess the bound $B_1$ using binary search instead of enumerating through all values in the range $[1, B]$. Moreover, we will assume that $L$ is polynomially bounded. This assumption will become clear when we use \sto with distance constraints to solve the buy-at-bulk problem in directed graphs.

Next, we show how to reduce the depth of our recursion at the cost of additional guessing. The idea is the same as in Section \ref{subsec:imp}: in one level of the new algorithm we guess all quantities in $s=\epsilon\cdot \log\log k$ levels of Algorithm~\ref{rec-greedy-dc}. As 
the number of subproblems generated at each level of recursion in Algorithm~\ref{rec-greedy-dc} is $2n2^d BL^2$,  each subproblem in the new algorithm generates $(2n2^d BL^2)^{2^s}$ many   subproblems.  
Applying the binary search idea (Section~\ref{subsec:qp}) on top of this results in bringing down the dependence on $B$ to $\log B$.\footnote{We do not see how to  reduce the linear dependence on the lengths $L$ to logarithmic in $L$.} So the overall running time is $(nL\log B)^{O(\log^{1+\epsilon} k)}$. This completes the proof of Theorem~\ref{thm-lc}.

\subsection{\sto with Deadlines}\label{subsec:sto-deadline}

For the second extension, along with the input to \sto, we have a length function $\ell: E \to \mathbb{Z}_{+}$, and deadlines $\{d_v\}_{v \in V}$. We are able to claim the reward of a vertex $v$ in arborescence $T$ only if $l_T(v) \leq d_v$. The goal of the problem is to find an out-directed arborescence $T^*$ rooted at $r^{*}$ maximizing $f(S(T^*))$ such that $c(T^*) \leq B$ where $S(T^*) = \{v \in V: l_{T^*}(v) \leq d_v \}$. We call this problem \sto with Deadlines. Note that $l_{T}(v)$ and $r^{*}$ are as defined in section \ref{subsec:sto-lc}.

The main algorithm is implemented by the procedure RG-DL$(r, k_r, Y, D, B, X, i)$. The parameters are the same as described in Section \ref{subsec:sto-lc}. We would like to point out the differences between \sto with length constraints and \sto with deadllines.
\begin{itemize}
\item First of all, we do not compute $f$ on all the vertices in the arborescence that is returned. If $T$ is the arborescence that is returned, then $f$ is computed on the set $S(T) = \{ v \in T: \l_{T}(v) \leq d_v\}$.
\item Second of all, there is no length bound on a feasible arborescence. Thus, we do not need to guess the total length $L_1$ and $L - L_1$ for the two arborescences rooted at $r$ and $v$ respectively, where $v$ is a separator vertex for the arborescence rooted $r$. Although, there is no length bound, we still need to guess the length from $r$ to $v$. Suppose $T^*$ is the optimal arborescence rooted at $r$, and $T_1^*$ and $T_2^*$ are its components rooted at $r$ and $v$ respectively. If we do not reach vertex $v$ from $r$ with a distance less than $l_{T^*}(v)$, then we will not be able to claim the objective claimed by $T_2^*$ due to the added deadlines on the vertices.
\end{itemize}

The recursive-greedy algorithm to solve this problem is  similar to Algorithm \ref{rec-greedy-dc} with the above two changes. The analysis for the runtime, and the approximation guarantee are also similar to the ones presented earlier. For brevity, we will omit it here, and state the final result. 

\begin{theorem}
There is an $O(\frac{\log k}{\log \log k})$-approximation algorithm that runs in quasi-polynomial time for the submodular tree orienteering problem with deadlines.
\end{theorem} 

\subsection{Single source Buy-at-Bulk} \label{subsec:bab}
Here we use the approximation algorithm for \sto with length constraints to obtain an approximation algorithm for the single source buy-at-bulk problem in directed graphs. In this problem, we are given a directed graph $G(V, E)$, a set of terminals $S$ and a source/root $r^*$. Moreover, each edge $e\in E$ is associated with a monotone concave cost function $g_e:\mathbb{R}_+ \rightarrow \R_+$. The goal is to route a unit of flow from $r^*$ to each terminal in $S$ while minimizing the total cost $\sum_{e\in E} g_e(x_e)$  where $x_e$ denotes the total flow through edge $e$. It is straightforward to show (using concavity) that the edges carrying non-zero flow must form an $r$-arborescence. We adopt an alternative representation of the buy-at-bulk problem (at the loss of a constant factor in approximation) as described in \cite{MMP08,CHKS10}. The input to the problem is now a directed multi-graph $G(V, E)$, a cost function $c: E \to \R_+$, a length function $\ell: E \to \R_+$, a set of terminals $S$, and a source $r$.  
The goal is to find an $r$-rooted arborescence $T$ that has a directed path to all terminals such that $\sum_{e \in T}c(e) + \sum_{v \in S}\ell_T(v)$ is minimized. Here too, the function $\ell_T(\cdot)$ denotes the length of the path from $r$ to $v$. 

\def\tl{\tilde{\ell}}
\def\tb{\widetilde{B}}
We follow a set covering approach as used in section \ref{sec:app}. 
We first guess an upper bound  $B$ on the optimal value, which implies  the same bound on the cost  $\sum_{e \in T}c(e)$ and total length $\sum_{v \in S}\ell_T(v)$ of the  optimal arborescence $T$. The guessed bound will be  guaranteed to be within a factor of $2$ of the optimal by using a binary search approach. Then we will iteratively run the algorithm for \sto with length constraints with cost and length bounds $B$, and a reward of one for all \textit{uncovered} terminal vertices. 
 Notice that the term $\sum_{v \in S} \ell_T(v)$ in the objective only takes into account the terminal vertices, and not all vertices in the arborescence. Algorithm \ref{rec-greedy-dc} can easily be modified to incorporate this change. Assuming that the bound $B$ is guessed correctly, the optimal value of each \sto instance solved above equals $k'$, the number of uncovered terminals.  As we use a $\rho={O}(\frac{\log k}{\log \log k})$ approximation for \sto with length constraints (Theorem~\ref{thm-lc}), the number of iterations before covering all terminals is at most $O(\rho\cdot \log k)$. 

Another (minor) issue is that   Theorem~\ref{thm-lc} requires polynomially-bounded lengths. 
In order to ensure this, we perform a standard scaling/rounding as follows. We first remove all edges $e\in E$ with $\ell_e>B$ as these will not be used in an optimal solution. Let $\widetilde{B}=B/n^4$ and $\tl_e = \lfloor \ell_e/\tb\rfloor$ for all $e\in E$. Note that the new lengths $\tl_e$ are integers between $0$ and $n^4$, so it is poly-bounded. We will run the algorithm from Theorem~\ref{thm-lc} on the instance with costs $c_e$, cost bound $B$, lengths $\tl_e$ and length bound $L=n^4$. Note that $\tb\cdot \tl_e \le \ell_e \le \tb\cdot \tl_e +\tb$ for all $e\in E$. It is clear that the optimal arborescence $T$ to the buy-at-bulk instance is still feasible to this \sto instance with the new length constraint. On the other hand, any arborescence $\widetilde{T}$ satisfying the new length constraint has total length  
$$\ell(\widetilde{T})\le \tb\cdot \tl(\widetilde{T}) + n^2\tb \le n^4 \tb + n^2 \tb \le (1+o(1))B, $$ where we use the fact that each path has at most $n$ edges and there are at most $n$ terminals (whose paths contribute to the objective). 

\begin{theorem}\label{thm:bab}
There is a quasi-polynomial time  ${O}(\frac{\log^2 k}{\log \log k})$-approximation algorithm for the single-source buy-at-bulk problem in directed graphs. 
\end{theorem}

\subsection{Priority Steiner tree}\label{subsec:priority}
This is a generalization of Steiner tree that has been used to model quality-of-service (QoS) considerations~\cite{CNS04}. 
In the {\em priority} Steiner tree problem, we are given a directed graph $G(V, E)$ with edge-costs $\{c_e:e\in E\}$, a set of terminals $S$ and a root $r^*$. There are $p$ priority levels, with $1$ denoting the lowest and $p$ denoting the highest priority levels. Each edge $e$ has a priority $\theta_e$ which denotes its QoS capability.  
Each terminal $t\in S$ also has a priority $\lambda_t$ which denotes its QoS requirement. The goal is to find a minimum cost $r^*$-arborescence where the $r^*-t$ path for each  terminal $t\in S$ has all  edges with priority at least $\lambda_t$.

Algorithm~\ref{rec-greedy-dc} can be easily extended to the ``maximum coverage'' version of priority Steiner tree, where we are given a bound $B$ on cost and want an arborescence that priority-connects the maximum number of terminals. We say that a terminal $t$ is {\em priority-connected} if each edge in the $r^*-t$ path has priority at least $\lambda_t$. A recursive subproblem here is given by parameters $r,p_r,Y,D,B,X,i$. We just point out the  differences from the recursive parameters in Algorithm~\ref{rec-greedy-dc}. There is no longer a length constraint, so the parameters $k_r$ and $L$ are not needed. Instead, we  track the minimum priority $p_r$ of any edge on the path from $r^*$ (original root) to $r$ (current root). We also need to maintain  priorities in the dictionary $D$, which now contains vertex-priority pairs for all vertices in the responsibility set $Y$: in particular, the returned arborescence must ensure that all edges on the $r-w$ path for any $w\in Y$ have priority at least $D(w)$. For the recursion, as before we guess the separator vertex $v$, responsibilities $S\sse Y$ and budget $B_1$ for the $r$-rooted subproblem. In addition, we guess the priority level $q$  of $v$ which is passed to the two subproblems as follows: $D(v)=q$ for the $r$-rooted subproblem and $p_v=q$ for the $v$-rooted subproblem. Finally, in the base case ($i=1$) we check that the edge $(r,v)$ to vertex $v\in Y$ has priority at least $D(v)$.  This leads to an 
${O}(\frac{\log k}{\log \log k})$-approximation algorithm in for the maximum-coverage version in $n^{O(\log^{1+\epsilon} k)}$ time. We omit the proof details as it is very similar to that for Algorithm~\ref{rec-greedy-dc}. 

Combined with the set-covering framework as before, we obtain:
\begin{theorem}\label{thm:priority}
There is a quasi-polynomial time  ${O}(\frac{\log^2 k}{\log \log k})$-approximation algorithm for the priority Steiner tree  problem in directed graphs.
\end{theorem}


\newcommand{\etalchar}[1]{$^{#1}$}

\appendix

\section*{Proof of Lemma~\ref{lem-alg2}}
\textit{Proof.} We prove the lemma by induction on $i$. For the base case, let $i=1$. Since $T^*$ is feasible for $i=1$,  $T^*$ is either empty or contains a single edge. If $|Y| = 0$, then we guess the base-case vertex and return the one that maximizes $f_X$ subject to the given budget: so $f_X(T) \geq f_X(T^*)$ in this case. If $|Y| = 1$, then $T^*$ has a single edge, say $(r, v)$. Our procedure here will return the  arborescence with $(r, v)$, and so $f_X(T) = f_X(T^*)$. Thus, in either case, we get $f_X(T) \geq f_X(T^*)$ which proves the base case.

Suppose that $i > 1$. Let $v$ be the  vertex in $T^*$ obtained from Fact~\ref{fact-vertex-sep} such that we can separate $T^*$ into two connected components: $T_1^*$ containing $r$ and $T_2^*= T^* \setminus T_1^*$, where  $\max(|V(T_1^*)|, |V(T_2^*)|) \leq \frac{2}{3}|V(T^*)|$. Note that $T^*_1$ is an $r$-rooted arborescence that contains $v$ and $T^*_2$ is a $v$-rooted arborescence. 
Let $Y_2 \subseteq Y \setminus \{v\}$ be those vertices of $Y\setminus v$ that are contained in $T_2^*$, and let  $Y_1 = Y \setminus Y_2$. Because $T^*$ contains $Y$, it is clear that $\{v\}\cup Y_1\cup Y_2\supseteq Y$.  Finally, 
let $c(T_1^*) = B_1^*$ and $c(T_2^*) = B_2^* \leq B - B_1^*$. Note also that  
 $|V(T^*) \setminus \{r\}| \leq (\frac{3}{2})^i$. By the property of the separator vertex $v$, $\max(|V(T_1^*)|, |V(T_2^*)|) \leq \frac{2}{3}|V(T^*)| \leq (\frac{3}{2})^{i-1} + \frac{2}{3}$. Excluding the root vertex in $T_1^*$ and $T_2^*$, the number of non-root vertices in either arborescence is $\leq (\frac{3}{2})^{i-1}$. Let $f_X(T_1^*) = U_1$. We set $B_1$ in the algorithm using a binary search approach. Since we iterate over all values in $[1, U]$, one of the guesses, say $u'$, is $= \lceil U_1/(i-1) \rceil$. By the induction hypothesis, the value of the arborescence returned by RG-QP$(r, Y_1 \cup \{v\} \setminus \{r\}, B_1^*, X, i-1) \geq U_1 / (i-1)$. Also notice that RG-QP$(r, Y, b, X, i-1)$ is an increasing function in the parameter $b$ (this allows us to use binary search to find $B_1$). Thus, the value $B_1 \leftarrow \min_b(\text{RG-QP}(r, Y_1 \cup \{v\} \setminus \{r\}, b, X, i-1) \geq  u')$ has the property that $B_1 \leq B_1^*$.
 
 We can thus claim that:
 \begin{equation}\label{eq:qp-left-rec}
 T_1^* \mbox{ is compatible with }(r, Y_1 \cup \{v\} \setminus \{r\}, B_1, X, i-1) \mbox{ and}     
 \end{equation}
 \begin{equation}\label{eq:qp-right-rec}
   T_2^*   \mbox{ is compatible with } (v, Y \setminus (Y_1 \cup \{v\}), B - B_1, X \cup V(T_1), i-1).
 \end{equation}
Now consider the call $\text{RG}(r, Y, B, X, i)$. 
Since we iteratively set every vertex to be the separator vertex, one of the guesses is $v$. Moreover, we iterate over all subsets $S \subseteq Y$, and thus some guess must set $S = Y_1$. From the above argument, one of the guesses $u \in [1, B]$ gives us $B_1 \leq B_1$. Thus, we see that one of the set of calls made is 
$$T_1 \leftarrow \text{RG}(r, Y_1 \cup \{v\} \setminus \{r\}, B_1, X, i-1) \mbox{ and }T_2 \leftarrow \text{RG}(v, Y \setminus (Y_1 \cup \{v\}), B - B_1, X \cup V(T_1), i-1)$$

We now argue that $T = T_1 \cup T_2$ has the property that $f_X(T) \geq f_X(T^*)/i$. By \eqref{eq:qp-left-rec} and  induction,  
\begin{equation} \label{eq-qp-t1}
f_X(T_1) \geq \frac{1}{i-1}f_X{(T_1^*)}
\end{equation}

Let $X' = X \cup V(T_1)$. Similarly, by \eqref{eq:qp-right-rec} and  induction, we have
\begin{equation} \label{eq-qp-t2}
f_{X'}(T_2) \geq \frac{1}{i-1}f_{X'}{(T_2^*)}
\end{equation} 
The rest of this proof is identical to the proof of Lemma \ref{lem-alg1}.
\hspace*{\fill} \qedsymbol

\end{document}